\documentclass[12pt,preprint]{aastex}

\usepackage{amssymb}
\usepackage{amsmath}
\usepackage{graphicx}

\newcommand{\f}    {\frac}

\newcommand{\epscr}  {\epsilon_{lr}}
\newcommand{\zetalr}  {\zeta_{lr}}
\newcommand{\omcp}  {\omega_{cp}}
\newcommand{\omcr}  {\omega_{cr}}

\newcommand{\vai}  {v_{\mbox{\scriptsize{Ai}}}}

\newcommand{\cms}{{\rm cm}\,{\rm s}^{-1}}

\newcommand{\s}{{\rm s}}

\newcommand{\GeV}{{\rm GeV}}
\newcommand{\GeVs}{{\rm GeV}\,{\rm s}^{-1}}
\newcommand{\GeVcm}{{\rm GeV}\,{\rm cm}^{-3}}

\newcommand{\bk}{{\mbox{\boldmath$k$}}}

\newcommand{\bp}{{\mbox{\boldmath$p$}}}

\newcommand{\bu}{{\mbox{\boldmath$u$}}}
\newcommand{\bvv}{{\mbox{\boldmath$v$}}}
\newcommand{\bx}{{\mbox{\boldmath$x$}}}

\newcommand{\bB}{{\mbox{\boldmath$B$}}}
\newcommand{\bD}{{\mbox{\boldmath$D$}}}
\newcommand{\bE}{{\mbox{\boldmath$E$}}}
\newcommand{\bF}{{\mbox{\boldmath$F$}}}

\newcommand{\bK}{{\mbox{\boldmath$K$}}}

\newcommand{\bnabla}{\mbox{\boldmath$\nabla$}}

\newcommand{\kpar}  {k_{\parallel}}
\newcommand{\ppar}  {p_{\parallel}}
\newcommand{\vpar}  {v_{\parallel}}
\newcommand{\kper}  {k_{\perp}}
\newcommand{\Dpar} {D_{\parallel}}
\newcommand{\Epar} {E_{\parallel}}

\newcommand{\Dper} {D_{\perp}}

\begin{document}

\title{Cosmic Ray History and its Implications for Galactic Magnetic Fields}

\author{Ellen G. Zweibel\altaffilmark{1}}
%\author{\today}
\altaffiltext{1}{JILA \& Department of Astrophysical \& Planetary Sciences,
 University of Colorado, Campus Box 440,
                 Boulder, CO 80309-0440, USA}

\shorttitle{Cosmic Rays and Magnetic Fields in Young Galaxies}
\shortauthors{Zweibel}

\begin{abstract}
There is evidence
that cosmic rays were present in galaxies
at moderately high redshift. This suggests that magnetic fields were also 
present. If cosmic rays and magnetic fields must always be close to 
equipartition, as they are to an order of magnitude within the local universe,
this would provide 
a powerful constraint on theories of the origin and evolution of
magnetic fields in galaxies. We evaluate the role
of magnetic fieldstrength in cosmic ray acceleration and confinement. We find
that the properties of small scale
hydromagnetic turbulence are fundamentally changed in the presence of cosmic
rays. As a result,
magnetic fields several orders of magnitude weaker than present
galactic fields can accelerate and retain a population of
relativistic cosmic rays, provided that
the fields are coherent over lengthscales
greater than a 
cosmic ray gyroradius. 
\end{abstract}
\keywords{cosmic rays --- galactic evolution---magnetic fields---MHD turbulence}

\clearpage

%%%%%%%%%%%%%%%%%%%%%%%%%%%%%%%%%%%%%%%%%%%%%%%%%%%%%%%%%%%%
%
% section{Introduction}
%
%%%%%%%%%%%%%%%%%%%%%%%%%%%%%%%%%%%%%%%%%%%%%%%%%%%%%%%%%%%%
\section{Introduction}

Superthermal particles, or cosmic rays, are a major constituent of
the interstellar medium in galaxies. They are dynamically coupled
to the thermal gas, and collectively
provide pressure support, can drive
outflows, and are thought to modify the structure of shocks.
Cosmic rays are the
dominant source of ionization, and a major source of heating, in regions of
high visual extinction. Spallation reactions 
between cosmic rays
and ambient interstellar nuclei are the primary mechanism for synthesizing
light elements.

The energy spectrum, directional anisotropy, and chemical 
composition of cosmic rays are measured directly through
\textit{in situ} observations on the Earth
and in the heliosphere. The
global properties of cosmic
rays in the Galaxy are probed through their
radio frequency synchrotron emission and $\gamma$-ray line and
continuum emission. A recent summary of the observations is given by
\citet{S02ii}.

Although 
synchrotron and $\gamma$-ray radiation from normal galaxies
have up to now been directly
detected only within the local universe, there
is evidence that cosmic rays existed in galaxies even at early times.
The light elements present in metal poor halo stars are thought to originate
by spallation reactions between energetic particles and the
material from which these stars formed
[\citet{DLL92}, \citet{G92}, \citet{D98}, \citet{B99} and references
therein]. Further evidence is provided by
the spectrum of the
diffuse $\gamma$-ray background, which is best fit by
a superposition
of active and normal galaxy $\gamma$-ray
spectra at moderate redshift [\citet{VF02}].

The
energy densities in the Galactic magnetic field
and cosmic rays are now roughly equal. If this
relationship is universal and necessary, then the presence of
cosmic rays in young galaxies implies the existence of
equipartition magnetic fields. This would be an important constraint on
theories of the
history of magnetic fields in galaxies, a subject on which evidence is
sparse.
At present, the only direct observations of magnetic fields associated with
young galaxies are the Faraday rotation measurements in damped Ly${\alpha}$
systems at redshift $z\le 1$ reported by \citet{OW95}.
These observations suggest magnetic fields of $\sim 1 \mu$G
strength which are spatially coherent over several kpc, similar to the fields
in the Milky Way and other spiral galaxies. Inferences based on cosmic rays
could probe galactic magnetic field evolution at earlier epochs, and,
we will see,  could be directly
relevant to conditions in star forming regions. This is significant because of
the important role magnetic fields can play in star formation and their
possible effect on the Initial Mass Function.

The leading theories of cosmic ray acceleration and confinement depend 
on the strength of the magnetic field. In this paper we use this
dependence to estimate the minimum magnetic field at which a population of
relativistic cosmic rays with energy density comparable to the present 
Galactic population
can be accelerated and confined. We find that a
field
several orders of magnitude weaker than the present Galactic
 field is sufficent for both acceleration and confinement, 
provided that the coherence length of the field exceeds
the cosmic ray gyroradius. 

The plan of this paper is as follows. In \S\ref{sec:early} we state our
assumptions about those
conditions in early galaxies which affect the acceleration and propagation of
cosmic rays.
In \S\ref{sec:review} we briefly review
the theory of energetic particle transport and its application to cosmic rays.
In \S\ref{sec:dmin} we derive constraints which
arise by assuming that the cosmic rays diffuse at the minimum possible
rate, which at a given fieldstrength gives the maximum acceleration and 
confinement. In \S
\ref{sec:highbeta} we discuss hydromagnetic
turbulence in a weak
magnetic field and find that it is drastically affected by even a small
population of cosmic rays.
Fluctuations grow rapidly in the presence of cosmic ray anisotropy, and, in
contrast to fluctuations driven by cosmic ray anisotropy in the present
Galactic environment, are not
Alfv\'enic in character. At small amplitudes, the damping rate is much lower
than the excitation rate. Therefore, we predict that the waves grow to
nonlinear amplitudes.
This may have implications for the nature of turbulence in
other environments in which the
magnetic fields is relatively weak and a population of energetic particles
is present.
Section \ref{sec:discussion} is a summary and conclusion. Most of the
technical material is in Appendices A - C.

We use
Gaussian cgs units throughout, except in expressing the energies of
cosmic rays, where we follow standard practice and use eV, and also in
certain formulae, in which we use
``natural" units such as parsecs or years.
%%%%%%%%%%%%%%%%%%%%%%%%%%%%%%%%%%%%%%%%%%%%%%%%%%%%%%%%%%%%
%
% section{Early}
%
%%%%%%%%%%%%%%%%%%%%%%%%%%%%%%%%%%%%%%%%%%%%%%%%%%%%%%%%%%%%

\section{Conditions in Early Galaxies\label{sec:early}}

We assume that the density and temperature of interstellar gas
in young galaxies are in the range found in contemporary galaxies.
At lower metallicity the cooling rate
is reduced, and the period in which supernova remnants are nonradiative is
extended. This moderately enhances the efficiency of cosmic ray acceleration. 
We assume that Type II supernovae are clustered, as they are
now, and may occur at a higher rate than  at present.
We consider the overall galactic size scale to be similar to
present galaxy sizes, or perhaps somewhat smaller. The values we choose are
appropriate to spheroidal systems with radii comparable to the present
thickness of galactic disks.

The most important assumptions concern the magnetic field.
A recent review of
the origin of cosmological magnetic fields is given by \citet{W02}, and
discussions of magnetic fields in galaxies are given in \citet{B96} and
\citet{ZH97}. Briefly, theories of the origin of galactic magnetic fields can
be classified as
top down or bottom up. In top down theories,
the fields arise through large scale coherent processes and have
large scale structure, although they are generally very weak. Examples of
top down theories include magnetogenesis during inflation \citep{TW88}, by
the operation of the Biermann battery in
cosmological shock fronts or other vortex 
structures [\citet{PS89}, \citet{K97}, \citet{DW00}], 
in cosmological
ionization fronts [\citet{SNC94}, \citet{GFZ00}], 
and through the manyfold expansion of plasma lobes
associated with active galactic nuclei [\citet{FL01}, \citet{K01}]. 
Although only the first of these theories produces fields
which are truly coherent over cosmological scales, all three produce fields
which have some coherence over galactic scales.

In bottom up theories, there are many independent,
small scale sources of field. The magnetized plasma from these sources
expands and diffuses to fill the interstellar medium,
eventually merging to create pervasive fields with largely random structure.
Examples of the sources include plerion supernova remnants and jets or winds
from
accretion disks surrounding stellar mass compact objects
[\citet{Z83}, \citet{R87}]. The 
magnetic field is inhomogeneously distributed until it 
has had time to 
diffuse.
For example, if magnetic fields were seeded by massive stars, there
could have been fields in OB associations before there was a large 
scale galactic field.

Both top down and bottom up theories predict magnetic
fields which are much weaker than present galactic fields, and must be
amplified by
several orders of magnitude or more to resemble
the fields of several $\mu$G 
which are observed in
galaxies. It is thought that amplification is due to the action of a
hydromagnetic dynamo, which stretches the field by
fluid motions while dissipating it at the resistive scale. Since resistive
effects typically operate at
an AU or less, while the coherence lengths of galactic
magnetic fields are observed to be several kpc or more, a longstanding
issue is how the fieldlines can be lengthened, within a fixed volume, without
the extensive folding and tangling that characterizes a
predominantly small scale field 
[see e.g. \citet{KA92} and \citet{S02}] 

For our purposes, the chief distinctions between top down and bottom up 
theories are that the former predict a large scale, spatially homogeneous
field while the latter predict a small scale, inhomogeneous field. Partly in
recognition of bottom up theories,
we will discuss cosmic ray acceleration by supernova
shocks and propagation in the general galactic field more or less independently
of one another.
%%%%%%%%%%%%%%%%%%%%%%%%%%%%%%%%%%%%%%%%%%%%%%%%%%%%%%%%%%%%
%
% section{Review}
%
%%%%%%%%%%%%%%%%%%%%%%%%%%%%%%%%%%%%%%%%%%%%%%%%%%%%%%%%%%%%

\section{Transport Theory with Application to Galactic Cosmic Rays\label{sec:review}}

There is a well developed
theory for interactions between energetic particles and low frequency
electromagnetic fluctuations $\bB_1$, $\bE_1$ superimposed on
a mean magnetic field $\bB_0$. Reviews are given by
\citet{J71}, \citet{W74},
\citet{T85}, and \citet{S02ii}. This theory has been reasonably successful in
explaining the $\sim 10^7$ yr confinement time of cosmic rays, their near
isotropy, and their energy density, the latter provided that approximately
10\% of supernova energy is converted to cosmic rays. In the interests of
keeping the paper self contained, we 
provide a brief treatment here.

A particle with velocity $\bvv$ and
cyclotron frequency $\omcr
\equiv ZeB_0/mc\gamma$
interacts primarily with the Fourier components of $\bB_1$ and $\bE_1$ which
satisfy the cyclotron resonance condition
\begin{equation}\label{equ:resonance}
\omega - \kpar\vpar \pm\omcr=0,
\end{equation}
where $\omega$ and $k$ are the frequency and wavenumber of the Fourier mode and
the subscript ``$\parallel$" denotes the projection along $\bB_0$.

If the spectrum of waves is incoherent, if a
particle interacts with a wave for one gyroperiod, and if the effect of any
single interaction is small, then the 
the wave-particle interaction 
produces diffusion in momentum space. Although
the particles diffuse in both the magnitude and direction of
their momentum $\bp$, the diffusion in angle is dominant as long as
the phase velocity of the resonant waves is much
less than the speed of light. The rate of angular 
scattering as a function of the variable $\mu\equiv\ppar/p$ is 
\begin{equation}\label{equ:nu}
\nu(\mu)\equiv\f{\pi}{4}\omcr\left(k_R{\mathcal{E}}(k_R)\right)^{-1},
\end{equation}
where
$\mathcal{E}$ is the power spectrum of the magnetic field fluctuations,
normalized to the large scale magnetic field, and $k_R$ is the value of $\kpar$
which satisfies eqn. (\ref{equ:resonance}).

The resonance condition eqn. (\ref{equ:resonance}) involves only the
parallel component of $\bk$. However, if $\kper\ge\kpar$, the 
oscillations of the wave electromagnetic fields over the gyrorbit diminish 
the response of the particle, and the scattering
frequency is reduced \citep{C00b}.

Locally, the
fluctuations define a frame with velocity $\bu_f$, which is the 
mean fluctuation velocity measured in the rest frame of the observer and
 weighted by the intensity spectrum. If the
fluctations propagate isotropically, or if the velocity
$\bu$ of the fluid is much larger than the relative velocities of the
fluctuations and the fluid, then $\bu_f$ can be approximated by $\bu$ itself. 
If the scattering time $\nu^{-1}$ and associated mean free path $\lambda_{\parallel}\equiv\mu v/\nu$ are short compared to the global lengthscales  and
timescales  of interest, the particles are well coupled to the fluctuation
frame, with a small amount of diffusion caused by scattering.
Under these conditions, the phase space distribution function
$f(\bx,\bp,t)$
of the cosmic rays is nearly isotropic with respect to $\bp$. The isotropic
part of $f$, averaged over the ensemble of fluctuations, is governed by a
transport equation which includes advection and
compression by the fluctuations, spatial diffusion 
with diffusion tensor $\bD$, and a source $S_0(\bx,\bp,t)$
\begin{equation}\label{equ:Ddiff}
\frac{\partial f_{0}}{\partial t} + \bu_f\cdot\f{\partial f_{0}}{\partial\bx} -
\f{1}{3}\left(\bnabla\cdot\bu_f\right)p\f{\partial f_{0}}{\partial p}
-\f{\partial}{\partial\bx}\cdot\bD\cdot\f{\partial f_0}{\partial\bx} =
S_{0}
\end{equation}
\citep{S75}.
Particle drifts caused by large scale curvature or gradients in $\bB_0$ can be
included explicitly in eqn. (\ref{equ:Ddiff}), in which case they appear in
the advection term \citep{FJO74}.

For our purposes, it suffices to consider a diagonal diffusion tensor with
components $\Dpar$ and $\Dper$. The parallel diffusivity is related to the
scattering frequency $\nu$ by
\begin{equation}\label{equ:dpar}
\Dpar\equiv\f{v^2}{4}\int_{-1}^{1}d\mu\f{1-\mu^2}{\nu(\mu)}
\end{equation}
According to eqns. (\ref{equ:nu}) and
(\ref{equ:dpar}), $\Dpar$ can be approximated by
\begin{equation}\label{equ:dparapprox}
\Dpar\sim\frac{1}{3}\frac{v^2}{\omega_c}\left(k_R{\mathcal{E}}(k_R)\right)^{-1}.
\end{equation}

The perpendicular diffusivity $\Dper$ is given approximately in
terms of $\Dpar$ by \citep{J87}
\begin{equation}\label{equ:dperp}
\Dper\sim\frac{\Dpar}{1+\left(k_R{\mathcal{E}}(k_R)\right)^{-2}}.
\end{equation}
Equation
(\ref{equ:dperp}) shows that cross field diffusion
is weak if the fluctuation amplitude is small,
but  becomes nearly isotropic if the fluctuations are
nonlinear ($k_R{\mathcal{E}}(k_R)\sim 1$).
According to eqn.
(\ref{equ:nu}), if the fluctuation amplitude is nonlinear, the scattering
frequency is approximately the cyclotron frequency. This is also the timescale
on which a particle interacts with a single wave.

Equation (\ref{equ:Ddiff}) is the foundation of the leading theories of
cosmic ray acceleration and confinement. It gives the confinement time $\tau_c$
to a region of size $L$ as
\begin{equation}\label{equ:tauc}
\tau_c\sim\f{L^2}{D},
\end{equation}
where $D$ is the largest component of the diffusion tensor,
generally $\Dpar$. Equation (\ref{equ:tauc}) is consistent with the observed
confinement time of Galactic cosmic rays if $k_R{\mathcal{E}}(k_R)\sim
10^{-6} - 10^{-7}$ \citep{C80}. Equation (\ref{equ:tauc}) is an upper limit in
the sense that it does not include transport due to migration of the fieldlines
themselves, which may be considerable if the lines of force have a large
stochastic component \citep{B90}. 

Equation (\ref{equ:Ddiff}) predicts that in a steady state in
which diffusion of cosmic rays is balanced by  sources, with little advection
or compression, the 
relationship between the cosmic ray energy density $U_{cr}$
and the power density of the source ${\mathcal{P}}_{cr}$ is given approximately
by
\begin{equation}\label{equ:Ucr}
U_{cr}\sim\tau_c{\mathcal{P}}_{cr}\sim{\mathcal{P}}_{cr}\f{L^2}{D}.
\end{equation}
In the Milky Way, $U_{cr}\sim 10^{-9}\GeVcm$, and $\tau_c$ is known from
the isotopic ratios of light elements to be $\sim 10^7$ yr. This implies that
${\mathcal{P}}_{cr}\sim 5\times 10^{-27}$ erg cm$^{-3}s^{-1}$, which is about
5\% of the power density in supernovae.

If the scattering is strong and the distribution function $f$ is nearly
isotropic, the cosmic rays can be treated as a fluid, and the dynamic and
energetic coupling of the cosmic ray and thermal fluids, which is mediated by
scattering, can be expressed in hydrodynamic form.
The rate of momentum transfer $\bF_{cr}$
takes the form of a pressure gradient force
\begin{equation}\label{equ:Fcr}
\bF_{cr}=-\bnabla P_{cr}
\end{equation}
acting on the
thermal fluid. The rate of energy transfer to the thermal fluid $\dot E_{cr}$
can be written as
\begin{equation}\label{equ:dotEcr}
\dot E_{cr}= -\gamma_{cr}(\bu_f-\bu)\cdot\bnabla
P_{cr}
\end{equation}
\citep{VDM84}, where $\gamma_{cr}$, the polytropic exponent of the
cosmic rays, is usually taken to be 4/3.
The rate of energy transfer is proportional to the
frictional force between the cosmic rays and thermal fluid.
The dynamical effects represented by eqns. (\ref{equ:Fcr}) and (\ref{equ:dotEcr}) are responsible for
hydrostatic support, acceleration of outflows, and
modification of shock structure in the thermal interstellar gas.

According to the leading theory of cosmic ray origin [original
papers by \citet{ALS77},
\citet{B78a}, \citet{B78b}, \citet{BO78}; reviews by \citet{BE87}, \citet{JE91}],
the particles are accelerated by
the first order Fermi process operating in strong
interstellar shocks driven by supernovae. Acceleration requires
hydromagnetic turbulence
both upstream and downstream of the shock. In a frame in which the shock is
steady, the upstream turbulence approaches the shock at nearly the shock
speed $u_S$ (the wave speed being negligible compared to the shock speed), and
the downstream turbulence recedes from the shock at speed $u_S/r$, where $r$ is
the compression ratio. Particles
which propagate upstream away from the
shock, are reflected back across the shock, and are then
scattered back upstream by the postshock turbulence gain an angle averaged net
momentum of $4m\gamma u_S(r-1)/3r$, much as if they were trapped between
converging walls.

The steady state spectrum of particles accelerated by a parallel shock is a
power law in momentum space, $f(p)\propto p^{-q}$, where the spectral index
$q$ is a function only of $r$: $q=3r/(r-1)$. The
power spectrum is realized
only within a range $p_{min} < p < p_{max}$. The lower limit $p_{min}
$ is the minimum momentum at which the conditions of
shock acceleration theory are
fulfilled: the particle speed $v$ must exceed the shock speed $u_S$ and
Coulomb losses must be insignificant.
The upper limit $p_{max}$
is the maximum momentum a particle can have reached
since the onset of acceleration. In \S\ref{sec:dmin}, we
will use the expression for $p_{max}$ derived by \citet{LC83}
to derive a lower limit on the magnetic fieldstrength such that shocks can
accelerate particles to relativistic energies.

The original theory of shock acceleration predicts the shape of the
energetic particle spectrum,
but not its normalization. The latter depends on the mechanisms by which
particles are injected into the shock. This so-called injection problem is
closely related to the issue of how the energetic particles modify the
structure of
the shock itself. 
We will parameterize the injection rate by an efficiency factor
representing the fraction of supernova energy which is converted to cosmic
rays. Estimates based on theory and/or numerical experiment suggest
that the efficiency is likely to be at least 10\% [\citet{EE84}, \citet{BE87}
\citet{JE91}, \citet{BE99}, \citet{MDV00},
\citet{E02}]

%%%%%%%%%%%%%%%%%%%%%%%%%%%%%%%%%%%%%%%%%%%%%%%%%%%%%%%%%%%%
%
% section{Constraints at the Minimum Diffusion Rate}
%
%%%%%%%%%%%%%%%%%%%%%%%%%%%%%%%%%%%%%%%%%%%%%%%%%%%%%%%%%%%%

\section{Constraints at the Minimum Diffusion Rate\label{sec:dmin}}

In this section, we assume that cosmic rays are scattered by
nonlinear
turbulence: $k_R{\mathcal{E}}(k_R)\sim 1$. According to eqns. (\ref{equ:dpar})
and (\ref{equ:dperp}),
the spatial diffusion coefficients $\Dpar$
and $\Dper$ are approximately
\begin{equation}\label{equ:dmin}
\Dpar\sim\Dper\sim\f{v^2}{3\omcr}\sim 3.1\times 10^{16}\f{\gamma\beta^2}{B}
\f{A}{Z},
\end{equation}
for particles with charge $Z$, atomic number $A$, speed $\beta c$, and Lorentz
factor $\gamma$.
Equation (\ref{equ:dmin})
corresponds to scattering
by one gyroradius $r_g$ per gyroorbit, which
is generally thought to be the minimum
rate at which cosmic rays can diffuse. We defer the question of whether such
a high level of turbulence can be maintained to \S\ref{sec:highbeta}. Here, we
merely find the minimum value of $B$ which is consistent with the
theories of cosmic ray
acceleration and propagation at maximum efficiency.

In this section and the following one, we focus on mildly relativistic cosmic
rays with energies of about 1 GeV/nucleon. This is slightly more than the
energy required for spallation reactions (a few hundred MeV/nucleon), and
the energy required to generate $\pi$ mesons from nuclear
collisions ($\sim$280 MeV/nucleon). That is, we focus on the energy range
required to create light elements and the $\gamma$-ray continuum resulting
from $\pi^{0}$ decay. We do not discuss the acceleration of electrons. They
are underabundant in cosmic rays, probably because they are injected at lower
efficiency, but the injection problem is beyond the scope of the present paper.

The acceleration
and propagation of the MeV cosmic rays responsible for ionization and heating
the ISM
is not, in our view, a plasma physics problem. The energy of these ions is
comparable to the energy per nucleon of supernova ejecta, and their range
against inelastic collisions
in neutral hydrogen is 10$^{20}$ - 10$^{21}$ cm$^{-2}$. Therefore, these
particles probably originate as supernova debris, and decay within their
first traversal of the
galaxy rather than being confined for times much longer than their transit
time. Interstellar gas has probably been heated and ionized by low energy
cosmic rays since the first generation of supernova explosions.

\subsection{Acceleration}\label{subsec:acceldmin}

\citet{LC83}
have derived the rate of particle acceleration by a strong parallel shock
when the fluctuation amplitude is unity. Their result can be written as
\begin{equation}\label{equ:egain}
\f{dE}{dt}=\f{3}{20}\f{ZeB}{c}u_S^2 = 1.5\times 10^{-18}ZBu_S^2\, \GeVs,
\end{equation}
where $B$ and $u_S$ are expressed in G
and $\cms$, respectively.

Equation (\ref{equ:egain}) can be integrated in time for any model of the
shock velocity $u_S(t)$. LC found that most of the energy gain occurs during
the blast phase, when the remnant is essentially undecelerated and $u_S$ is
constant, and the energy conserving Sedov phase, when $u_S\propto t^{-3/5}$.
The duration $t_b$ of the blast phase is
\begin{equation}\label{equ:tb}
t_b=6.2\times 10^9\f{M_e^{5/6}}{n^{1/3}E_{51}^{1/2}}\s,
\end{equation}
where $M_e$ is the mass of the ejecta in $M_{\odot}$, $n$ is the ambient
interstellar number density in cm$^{-3}$, and $E_{51}$ is the energy released
in the supernova explosion, in units of 10$^{51}$ erg. From eqn.
(\ref{equ:egain}),
\begin{equation}\label{equ:Eb}
E(t_b)=9.1\times 10^9\f{E_{51}^{1/2}}{M_e^{1/6}n^{1/3}}ZB\,\GeV.
\end{equation}
The energy gained between the beginning of the Sedov phase and its termination
by radiative cooling at time $t_r$ can be written in terms of the shock radii
$R_b$ and $R_r$ at times $t_b$ and $t_r$
\begin{equation}\label{equ:Es}
\Delta E_S = 1.3\times 10^{10}\f{E_{51}^{1/2}ZB}{M_e^{1/6}n^{1/3}}
\left(1-\left(\f{R_b}{R_r}\right)^{1/2}\right)\,\GeV.
\end{equation}
If the metallicity of the ambient medium is low, $R_r$ will be somewhat larger
than its contemporary value, somewhat increasing the energy gain. In any case,
Equations (\ref{equ:Eb}) and (\ref{equ:Es}) show that in order to accelerate
protons to relativistic energies, $E\sim \GeV$, $B$ must be at least $\sim
10^{-10}$G.

Since Type II supernovae are observed to be clustered, we also
consider a shock driven by expansion of a superbubble driven by a constant
(radiative plus mechanical) luminosity $L_0$
[\citet{SS95}]. In this case, $u_S\propto t^{-2/5}$ and the
bubble radius $R_S\propto t^{3/5}$. The particle energy
$E_{sb}$ increases steadily with time, and can be written in terms of the
bubble radius $R_S$ as
\begin{equation}\label{equ:Esb}
E_{sb}=8.2\times 10^8\left(\f{L_{38}R_{pc}}{n}\right)^{1/3}ZB\,\GeV,
\end{equation}
where $L_{38}$ and $R_{pc}$ are $L_0$ in units of 10$^{38}$erg s$^{-1}$ and
the bubble radius in pc, respectively. Equation (\ref{equ:Esb}) shows that
particle acceleration by superbubbles is generally weaker than acceleration
driven by single supernovae. A similar conclusion was reached by \citet{BF92} on
the basis of a somewhat different acceleration model.

If the upstream magnetic field lies nearly in the plane of the shock, instead
of nearly perpendicular to it, particles drift along the shock face while
scattering back and forth across it. The maximum energy
$E_d$ which can be acquired by drifting a distance $R$ is of order \citep{J87}
\begin{equation}\label{equ:Ed}
E_d\sim \f{ZeB}{c}u_SR=30ZBu_SR_{pc}\GeV.
\end{equation}
If $u_S\sim 10^9\cms$, and $R\sim$ 1 pc, $B$ must again be about 10$^{-10}$G
in order to accelerate protons to 1 GeV.

The requirement that the particle gyroradius must be less than the radius of
curvature of the shock also imposes a lower limit on the magnetic
fieldstrength. However, the gyroradius of a GeV proton in a 10$^{-10}$G
magnetic field is $\sim$0.01 pc, so the constraint arising from the
acceleration rate is more severe.

\subsection{Confinement}\label{subsec:confinemin}

The constraint that
the cosmic ray gyroradius must be less than the typical
magnetic lengthscale $L^B$ in
the galaxy is thought to set an upper bound on the energies
of cosmic ray nuclei in contemporary galaxies, and has led to the inference
that the highest energy cosmic rays are extragalactic in origin.
This limit arises because if $r_g > 
L^B$, the cross field drift velocity becomes comparable to the particle
velocity itself.  
The constraint derived from setting $r_g=L^B$ is
\begin{equation}\label{equ:rg}
B > 10^{-15}\f{A}{Z}\f{\sqrt{\gamma^2 -1}}{L^B_{kpc}},
\end{equation}
where 
$L^B$ is expressed in kpc.

The limit on $B$ derived from eqn. (\ref{equ:rg})
 is well below the value derived from the
acceleration time [eqns. (\ref{equ:Eb}), (\ref{equ:Es}), (\ref{equ:Esb}), or
(\ref{equ:Ed})], provided that $L^B$ is comparable to the scale of the galaxy
itself. However, it is a nontrivial constraint on those dynamo theories which
predict that much of the power
in the magnetic field is on small scales,
especially when the field is still very weak [\citet{KA92}, \citet{S02}]. If 
the power spectrum in the magnetic field peaked at the Ohmic scale, or even the
viscous scale, a much stronger field would be required to confine cosmic rays.

If we require that cosmic rays are confined 
within superbubbles long enough to produce light elements by
spallation reactions, the resulting condition on $B$ is more stringent
\footnote{Unraveling the history of light elements from measurements of their
relative abundances in the oldest stars is a complex matter which may require
more than one type of spallation source; see \citet{PD99}, \citet{PD00},
\citet{SI02}. We provide only a rough estimate here.}.
 Using eqns. (\ref{equ:dmin}) and
(\ref{equ:tauc}), the resulting constraint on $B$ for particles with Lorentz
factor $\gamma$ and speed $v=\beta c$ is
\begin{equation}\label{equ:bubbleconf}
 B > 1.0\times 10^{-7}\f{A}{Z}\f{\gamma\beta^2 \tau_6}{R_{pc}^2}G,
\end{equation}
where $\tau_6$ is $\tau_c$ expressed in units of 10$^6$ yr. Equation
(\ref{equ:bubbleconf}) shows that a field of 10$^{-11}$ G is sufficient to
confine GeV cosmic rays for 10$^6$ yr in a 100 pc superbubble, provided that
the turbulence is strong. This is comparable to, but slightly weaker than, the
field required for acceleration of such cosmic rays.

Equations (\ref{equ:Ucr}) and (\ref{equ:dmin}) lead to an expression for $B$
in terms of the energy density in cosmic rays $U_{cr}$ and the power density
of the cosmic ray source ${\mathcal{P}_{cr}}$, which we write as a fraction
$\epsilon$ of the supernova power density ${\mathcal{P}_{sn}}$. Taking
$\gamma\beta^2=1$ as average, setting $A=Z=1$ because most cosmic rays are
protons, normalizing $U_{cr}$ and
$P_{sn}$ by their contemporary values: $U_{cr}\equiv 10^{-9} u_{cr}\GeVcm$,
${\mathcal{P}_{cr}}
\equiv 10^{-22}p_{cr}
\epsilon\GeVcm\s$, and expressing the confinement lengthscale
$L$ in kpc gives
\begin{equation}\label{equ:galaxyB}
B= 3.3\times 10^{-14}\f{u_{cr}}{\epsilon L_{kpc}^2p_{cr}}G.
\end{equation}

Nonlinear models of cosmic ray acceleration in shocks suggest that $\epsilon\ge
$ 0.1. If all the other parameters take their fiducial values, a field of about
10$^{-12}$ - 10$^{-13}$G 
would sustain a cosmic ray energy density comparable to that
found in contemporary galaxies. An even lower fieldstrength would have
sufficed if the supernova rate was much higher at these early times than 
it is now.

%%%%%%%%%%%%%%%%%%%%%%%%%%%%%%%%%%%%%%%%%%%%%%%%%%%%%%%%%%%%
%
% section{Wave}
%
%%%%%%%%%%%%%%%%%%%%%%%%%%%%%%%%%%%%%%%%%%%%%%%%%%%%%%%%%%%%
\section{Wave Propagation, Excitation, and Damping in Weak Fields}
\label{sec:highbeta}

In this section we consider the propagation, excitation, and damping of
hydromagnetic fluctuations at very low magnetic fieldstrength, and attempt to
estimate whether early galaxies could have sustained a nonlinear level of
magnetic fluctuations.

At the magnetic fieldstrengths considered here, the
frequencies of Alfv\'en
waves at the  resonant wavenumbers are extremely low.
The Alfv\'en speed $\vai$ defined with respect to the ionized mass density is $
\vai\equiv B/(4\pi m_in_i)^{1/2}=2.2\times 10^{11}B/n_i^{1/2}$cm s$^{-1}$.
Motivated by the resonance condition eqn. (\ref{equ:resonance}), we define a
fundamental wavenumber $k_0$ in terms of the nonrelativistic proton
gyrofrequency $\omcp$
\begin{equation}\label{equ:k0}
k_0\equiv\f{\omcp}{c}=3.5\times 10^{-7}B\:{\mathrm{cm}}^{-1}.
\end{equation}
The Alfv\'en wave frequency is
\begin{equation}\label{equ:omegaest}
\omega\sim k_0\vai=7.3\times 10^4\f{B^2}{{n_i}^{1/2}}\:{\mathrm{s}}^{-1}.
\end{equation}
For example, if $B\sim 10^{-10}G$ and $n_i\sim 1$, the wave period is more than
10$^8$ yr.

The implication of the very long
Alfv\'en wave period is that the waves are
strongly affected by cosmic rays. Under
contemporary interstellar conditions, cosmic
rays influence hydromagnetic waves only through 
the resonant interaction, described
by eqn. (\ref{equ:resonance}). This means that only the
imaginary part of
the wave frequency is affected by the cosmic rays.
 But when the magnetic field is
weak, the electromagnetic response of the plasma is so feeble that even a low
density of cosmic rays has a significant effect on both the real and imaginary
parts of $\omega$. A similar situation prevails if a very
high flux of cosmic rays is present [\citet{KZ75}, \citet{Z79}].

At low magnetic fieldstrength,
Alfv\'en waves are also affected by the thermal plasma. 
\citet{FK79} have shown that the 
the large thermal ion gyroradius causes the ions to deviate from the
$\bE\times\bB$ drift which nearly exactly characterizes their motion in a cold
plasma, leading to modifications to the Alv\'en wave dispersion relation. 
Thermal plasma also causes efficient damping outside a
small cone of propagation angles centered on $\bB_0$.
It turns out that with the lower bounds on $B$ imposed in the previous section,
the thermal modifications to the real part of $\omega$ are not as important
as the cosmic ray modifications. We justify this statement below. 
Collisionless damping is potentially important, however, and we consider it in 
\S\ref{subsec:landau}.
Because the collisionless damping increases with the propagation angle
$\theta$ as $\theta^2$,
 while the excitation rate due to cosmic ray streaming is relatively
insensitive to $\theta$
for small $\theta$, we consider only excitation of waves with $\theta = 0$.

Much of the interstellar gas in early galaxies may have been only weakly
ionized, leading to damping of the waves by ion-neutral friction.
We consider weakly ionized gases in \S\ref{subsec:friction}.

In \S\ref{sec:early} we mentioned the possibility that young galaxies do not
have large scale magnetic fields. In discussing wave propagation, we tacitly
assume that there is a well ordered field on a scale at least several times
larger than the wavelength of the resonant wave; $\lambda\sim 6\times 10^{-12}
B^{-1}$ pc. Thus, a field of 10$^{-12}$G would have to be coherent over tens
of pc in order to support waves. This is much less than the current coherence
length of the Galactic magnetic field, which is at least a few kpc. However,
the possibility remains that even this degree of coherence is lacking. In this
case, it seems unlikely that the particles can be confined at all; see eqn.
(\ref{equ:rg}).

\subsection{Dispersion relation with cosmic rays}\label{subsec:dispersion}

We assume the cosmic rays are protons. As 
the presumptive dominant species, they control collective effects. The
dispersion relation for hydromagnetic waves in a medium with cold protons and
cosmic ray protons drifting relative to each other with speed $v_D$, and with
electrons cancelling the charge and current of each population of protons, is
derived in Appendix A [eqn. (\ref{equ:dr2})].
The dispersion relation is modified by cosmic rays
because of their drift and by the same mechanism that
modifies the dispersion relation in a hot plasma: the electron motion in the
wave
is nearly the $\bE\times\bB$ drift, but the cosmic rays deviate from the
$\bE\times\bB$ drift because of their large gyroradii. Therefore, the currents
associated with perturbations of
the two species do not come close to cancelling, and even at
low density the cosmic rays have a strong effect on the waves.

In the rest frame of the thermal plasma, the dispersion relation is
\begin{equation}\label{equ:dispreli}
\omega^2 + \omcp\f{n_{cr}}{n_i}\zetalr(k)\left(\omega-kv_D\right)-k^2\vai^2=0.
\end{equation}
The complex functions $\zetalr(k)$ (the subscript denotes the sign of circular
polarization)  are defined in Appendix A and plotted for a
sample cosmic ray distribution function in
Figure \ref{fig:zeta}. The real and imaginary parts of $\zeta$ represent
the contributions of
the nonresonant and resonant cosmic rays, respectively, and are of similar
magnitude for cosmic rays near the mean momentum of the distribution. Because
the $\zetalr$ cannot in general be easily evaluated, we treat $\zetalr$ as a
parameter of order unity in what follows, except in computing the group
velocity [see eqn. (\ref{equ:vg})].
If $v_D=\vai$, $\omega$ is real, and the waves are
marginally stable.

It is convenient to scale the wavenumber by
\begin{equation}\label{equ:kscale}
k=yk_0,
\end{equation}
where $k_0$ is defined in eqn. (\ref{equ:k0}),
and to absorb the ratio of cosmic ray to thermal plasma density into the $\zetalr$
by introducing 
\begin{equation}\label{equ:epscrdef}
\epscr\equiv\zetalr \f{n_{cr}}{n_i}.
\end{equation}

With these
definitions, the
general solution of eqn. (\ref{equ:dispreli}) is
\begin{equation}\label{equ:ionizedroots}
\f{\omega}{\omega_{cp}}=\f{1}{2}\left[-\epscr\pm\left(\epscr^2+4y^2
\f{\vai^2}{c^2}+4\epscr y\f{v_D}{c}\right)^{1/2}\right].
\end{equation}

The properties of eqn. (\ref{equ:ionizedroots}) depends on the
parameter
\begin{equation}\label{equ:Xdef}
X\equiv\f{\epscr}{y}\f{c}{\vai}\f{v_D}{\vai}
\sim 1.5\times 10^{-10}\left(\f{10^9n_{cr}\zetalr}{yn_i^{1/2}B}\right)\left(
\f{v_D}{\vai}
\right)
\end{equation}
which is the ratio of the third term under the radical to the second term.
This parameter has a more physical interpretation: if we assume that most of
the cosmic ray pressure is provided by mildly relativistic particles ($\gamma
\sim 1$) and use eqns. (\ref{equ:epscrdef}) and (\ref{equ:Xdef}), then the
ratio of the pressure in the mean magnetic field to the cosmic ray pressure
can be written approximately as
\begin{equation}\label{equ:Xphys}
\f{B^2}{8\pi P_{cr}}\sim\f{1}{2 X}\f{v_D}{c}.
\end{equation}
Equation (\ref{equ:Xphys}) shows that if $X$ is large, the magnetic field
must be far below equipartition with the cosmic rays.
 
Under contemporary interstellar conditions, $X\ll 1$ unless the cosmic ray flux
is much larger than average \citep{Z79},
and eqn. (\ref{equ:ionizedroots}) reduces to
\begin{equation}\label{equ:rootslimitold}
\f{\omega}{\omega_{cp}}=\pm y\f{\vai}{c} + \f{\epscr}{2}\left(1\pm\f{v_D}{\vai}\right).
\end{equation}
Equation (\ref{equ:rootslimitold}) agrees with the expression for the growth
rate given in \citet{KC71}.

The drift velocity $v_D$, diffusion coefficient $D$, and cosmic ray gradient
lengthscale $L^c$ are related by $v_D\sim D/L^c$. Setting $D$ equal to is
minimum value [eqn. (\ref{equ:dmin})] gives the minimum possible value of
$v_D/\vai$
\begin{equation}\label{equ:driftestimate}
\f{v_D}{\vai}\sim\f{\lambda}{L^c}\f{c}{\vai}\sim 4.7\times 10^{-17}
\f{n_i^{1/2}}{L^c_{kpc}B^2},
\end{equation}
where in the last step $L^c$ is expressed in kpc. At the fieldstrengths of 
interest here, $B\le 10^{-10}$G,
the anisotropy produced by cosmic ray sources drives the instability far
above threshold. Under these conditions,
the solution of eqn. (\ref{equ:ionizedroots})
is approximately
\begin{equation}\label{equ:rootslimitnew}
\f{\omega}{\omega_{cp}}=\pm\left(\epscr y\f{v_D}{c}\right)^{1/2}.
\end{equation}
The real and imaginary parts of eqn. (\ref{equ:rootslimitnew}) are comparable
in size, and the wave periods are much shorter than those implied by eqn.
(\ref{equ:omegaest}). Evaluating eqn. (\ref{equ:rootslimitnew}) numerically
gives
\begin{equation}\label{equ:rootsnumerical}
\omega=3.2\times 10^{-1}\f{B}{n_i^{1/2}}\left(\f{v_D}{c}\right)^{1/2}
\left(10^9 n_{cr}y\zetalr\right)^{1/2}.
\end{equation}

Although the turbulence required for diffusive shock acceleration might
always be present in the ambient medium, it is of interest to investigate the
circumstances under which the streaming instability grows quickly enough for
the cosmic rays to confine themselves. A rough estimate of $n_{cr}/n_i$ follows
from the efficiency assumed for shock acceleration and the ratio of the mean
cosmic ray energy to the postshock thermal energy: if 10\% of the
shock energy is converted to cosmic rays and the energy ratio is 10$^5$, then
$n_{cr}/n_i\sim 10^{-6}$. If $v_D\sim u_S\sim c/30$, eqn. (\ref{equ:rootsnumerical}) gives $\omega\sim 1.9 (y\zetalr)^{1/2}B$. If $B\sim 10^{-10}$ G, which is
approximately the limit implied by eqns. (\ref{equ:Eb}), (\ref{equ:Es}), and
(\ref{equ:Ed}), the growth time of the waves is a few hundred years. Thus, it
is not implausible that cosmic rays in the vicinity of a strong shock can trap
themselves.

In order to investigate confinement on the galactic scale, we express $v_D$
in terms of the confinement time $\tau_7$, the confinement time
$\tau_c$ in units of 
10$^7$ yr, and $L_{kpc}$, the confinement lengthscale in units of kpc.
If $\epscr\sim 10^{-9}$, $\omega\tau_c\sim 10^{12}B$, which 
exceeds unity as long as $B > 10^{-12}$G. This is comparable to, although
slightly larger than, the field required to maintain cosmic rays at their
present energy density [eqn. (\ref{equ:galaxyB})].

Finally, we return to the effect of thermal plasma on the wave frequency. 
According to the results of
\citet{FK79},
the ion gyroradius correction is represented by adding the term
\begin{equation}\label{equ:thermcorr}
\pm\omega\omega_{cp}\f{y^2}{2}\f{v_i^2}{c^2}
\end{equation}
to eqn. (\ref{equ:dispreli}),
where the $\pm$ signs again represent left and right circular polarization,
respectively, and $v_i^2\equiv 2k_BT/m_i = 1.6\times 10^8 T$. The
relative effect of
this thermal
term on the dispersion relation is measured by the quantity
\begin{equation}\label{equ:compare}
\f{y^3}{16\epscr}\left(\f{v_i}{c}\right)^4\f{c}{v_D}\sim 3.4\times
10^{-26}T^2\left(\f{n_i}{n_{cr}}\right)\left(\f{c}{v_D}\right).
\end{equation}
The right hand side of eqn. (\ref{equ:compare})
is typically much less than unity, so we neglect thermal corrections to $Re(\omega)$.

\subsection{Landau damping}\label{subsec:landau}

Unless
the angle $\theta$ between the wave vector and $\bB_0$ is exactly zero, the
wave electric field has a component $\Epar$ parallel to $B_0$. Thermal ions
which satisfy the  Landau
resonance condition $\omega=\kpar\vpar$ are steadily
accelerated by $\Epar$ and absorb energy from the wave. This process is
known as Landau damping, and is the
primary linear damping mechanism in fully ionized regions.

As we argue below, the condition $\theta\equiv 0$ is unlikely to be fulfilled
because of inhomogeneity in the background density and magnetic field.
Therefore, $\theta\ne 0$ is the generic case, and we must consider damping.

\citet{FK79}
evaluated the rate of Landau damping in a hot thermal plasma, with
$v_i^2/\vai^2\gg 1$. We show in Appendix B that their result is unchanged by
the presence of cosmic rays. The damping rate $\Gamma_L$
is given to within a factor
of order unity by
\begin{equation}\label{equ:landau}
\Gamma_L=\f{\sqrt{\pi}}{2}\kpar v_i\theta^2,
\end{equation}
 where we have approximated $\tan{\theta}$ by $\theta$
[see eqn. (\ref{equ:damping1})] within the range of
interest. 

We calculate the wave damping time in two ways.
First, we assume that
the waves are excited at $\theta=0$, but $\theta$ grows
as the waves propagate through the inhomogeneous background. We
define the Landau damping time $t_L$ by the condition
\begin{equation}\label{equ:tL}
\int_{0}^{t_L} \Gamma_Ldt=1,
\end{equation}
where $t$ is the time along a ray path.

Let $L^{b}$ denote the gradient lengthscale of the background, and $v_g$ the group
velocity of the wave; $v_g\equiv d\omega/dk$. The
evolution equation for $\theta$ is
\begin{equation}\label{equ:dtheta}
\f{d\theta}{dt}=\f{v_g}{L^b},
\end{equation}
with the solution $\theta=v_gt/L$ starting from zero initial condition. The 
solution of eqn. (\ref{equ:tL}) is 
\begin{equation}\label{equ:tLsol}
t_L=\left(\f{6}{\sqrt{\pi}\kpar v_i}\f{(L^{b})^2}{v_g^2}\right)^{1/3}.
\end{equation}
According to
eqns. (\ref{equ:rootslimitnew}) and (\ref{equ:epscrdef}), 
\begin{equation}\label{equ:vg}
v_g = \left(\chi_{lr}\f{n_{cr}}{n_i}\f{\omega_{cp}v_D}{4\kpar}\right)^{1/2},
\end{equation}
where $\chi_{lr}^{1/2}\equiv Re[\zetalr^{-1/2}d(k\zetalr)/dk]$.
Using eqn. (\ref{equ:vg}), we rewrite eqn. (\ref{equ:tLsol}) as
\begin{equation}\label{equ:tLsol2}
t_L=\left(\f{24(L^{b})^2}{\sqrt{\pi}v_Dv_i\chi_{lr}\omega_{cp}}\right)^{1/3}
\sim 3.0\times 10^{9}
\left(\f{(L^{b}_{pc})^2n_i}{10^{9}n_{cr}\chi_{lr}T^{1/2}B}\f{c}{v_D}\right)^{1/3}\:s,
\end{equation}
where $L^{b}_{pc}$ is $L^b$ expressed in pc.

Damping and excitation balance when $\omega t_L\sim 1$. Using eqns.
(\ref{equ:rootsnumerical}) and (\ref{equ:tLsol2}) yields the condition
\begin{equation}\label{equ:balance1}
1.1\times 10^9\left( B
L^{b}_{pc}\right)^{2/3}\f{\left(\zetalr y\right)^{1/2}}{\chi_{lr}^{1/3}}
\left(\f{10^9 n_{cr}}{n_iT}
\f{v_D}{c}\right)^{1/6}\sim 1.
\end{equation}
Evaluating the right hand side of
eqn. (\ref{equ:balance1}) with $v_D$ set equal to its minimum value given 
in eqn. (\ref{equ:driftestimate}) yields
\begin{equation}\label{equ:growth}
\omega t_L\vert_{min}= 
 2.9\times 10^{6}B^{1/2}\f{(L^{b}_{pc})^{2/3}}{(L^{c}_{kpc})^{1/6}}
\left(\f{10^9n_{cr}\zetalr}{n_iT}\right)^{1/6}.
\end{equation}
If $\omega t_L\vert_{min}$ exceeds unity, Landau damping is always weaker than
excitation by cosmic ray anisotropy. This is
likely to be the case for $B\ge 10^{-11} - 10^{-10}$G. Presumably,
the wave amplitude is limited to $B_1/B_0\sim 1$ by a nonlinear mechanism. 

As an alternative method of calculating the damping, we suppose
that the initial condition $\theta = 0$ can never be realized
in an inhomogeneous system. If instead of solving eqn. (\ref{equ:dtheta})
 we simply set
$\theta=(\kpar L)^{-1}$ and define the damping time $t_{L}^{\prime}$ as
$\Gamma_L^{-1}$, with $\Gamma_L$ taken from eqn. (\ref{equ:landau}), we recover
the condition on $v_D$ given by eqn. (\ref{equ:balance1}) with the factor of
1.1. reduced by about a factor of 2.
The insensitivity of the damping rate to the initial
conditions justifies our approximate treatment of the problem, and reinforces
the conclusion that Landau damping by the thermal plasma is weak compared with
excitation by cosmic rays.

\subsection{Ion-neutral damping}\label{subsec:friction}

Large volumes of the interstellar gas in young galaxies may be
only weakly ionized. In such regions, hydromagnetic waves are damped by
ion-neutral friction. Under present conditions, frictional damping
is so strong that cosmic ray stream through H I regions 
virtually  without scattering \citep{KC71}.

The rate $\Gamma_{in}$ at which Alfv\'en waves are damped by
friction is given in \citet{KP69}. The modification of the waves by cosmic rays
significantly reduces the damping rate.
As shown in Appendix C, most of the energy in the waves
is in the cosmic rays themselves, with a smaller fraction in electromagnetic
fields, and the energy carried by the thermal gas a distant third. 
Since the ions
are coupled to the wave by electromagnetic forces, while
collisions with the neutrals directly tap only the small amount of energy
carried by the ions, the overall effect of ion-neutral friction is
doubly weakened.

The damping rate is calculated by an energy method in Appendix C. 
The result can be written in terms of the ion-neutral collision
frequency $\nu_{in}\equiv\rho_n\langle\sigma v\rangle/(m_i+m_n)$ and the phase
velocity $v_{\phi}\equiv\omega/k$ as
\begin{equation}\label{equ:friction}
\Gamma_{in}=\nu_{in}\f{\vai^4}{v_{\phi}^2\left(v_{\phi}^2+\vai^2\right)}.
\end{equation}
We assumed in deriving eqn. (\ref{equ:friction}) that $\omega > \nu_{in}$.
In the parameter regime of interest, this appears to be well justified.

If the waves are Alfv\`en waves, $v_{\phi}=\vai$, and eqn. (\ref{equ:friction})
reduces to $\nu_{in}/2$. This is the usual expression for the
damping rate of high frequency
Alfv\'en waves in a
weakly ionized gas \citep{KP69}. According to eqns. (\ref{equ:Xdef}) and
(\ref{equ:rootslimitnew}),
\begin{equation}\label{equ:vphivai}
\f{v_{\phi}}{\vai}\sim X^{1/2}
\gg 1
\end{equation}
in the regime of interest. Not only is the frictional damping rate 
a factor of $X^2$ lower than it is for Alfv\`en waves, it actually
decreases as $v_D/c$ increases. Therefore, the final streaming rate cannot be
determined by balancing ion-neutral friction against excitation by cosmic
ray anisotropy, and the ionization state of the medium has
little relevance to cosmic ray acceleration or confinement.

%%%%%%%%%%%%%%%%%%%%%%%%%%%%%%%%%%%%%%%%%%%%%%%%%%%%%%%%%%%%
%
% section{Discussion}
%
%%%%%%%%%%%%%%%%%%%%%%%%%%%%%%%%%%%%%%%%%%%%%%%%%%%%%%%%%%%%

\section{Summary and Conclusions\label{sec:discussion}}

The light elements present in the oldest Galactic halo stars, and the spectrum
of the diffuse $\gamma$-ray background, are indirect evidence that cosmic rays
were present in young galaxies. Cosmic rays cannot be accelerated or confined
without some level of magnetization. In this paper, we investigated just how
strong a magnetic field is required, and what the presence of cosmic rays
implies about the evolution of galactic magnetic fields.

Cosmic rays can be confined in appreciable numbers only if the
magnetic coherence length $L^B$ is larger than their gyroradii.
This is the standard argument for the
extragalactic origin of the cosmic rays with energies above $\sim 10^9$ GeV.
Generalizing the argument to early times leads
to a lower limit on the product $BL^B$ [eqn. (\ref{equ:rg})]. If
$L^B$ were as large as a kpc, typical of the gradient lengthscale in the galaxy
itself, then $B$ must have exceeded 10$^{-15}$G if GeV protons were confined.
If the magnetic energy density was
concentrated at a much smaller scale such as the 
the Ohmic dissipation length,
cosmic rays probably could not have been
confined at all.

Under contemporary conditions, cosmic rays are scattered by gyroresonant 
interactions with small amplitude ($B_1/B_0\sim 10^{-3} - 10^{-4}$) 
hydromagnetic turbulence
[eqn. (\ref{equ:resonance})], and propagate diffusively. Although the
turbulence can in principle be excited by the streaming anisotropy of the 
cosmic rays themselves, it is also strongly damped, and it appears that 
supplemental sources of turbulence are required\footnote{If the turbulence
is highly anisotropic, so that the efficiency of scattering is reduced, then
processes other than gyroresonant scattering,
such as magnetic mirroring, may be necessary to explain
confinement \citep{C00a}}. The excitation of waves by cosmic rays is, however,
important in the vicinity of strong shock waves, where it is believed that the
bulk of galactic cosmic rays are accelerated.

If cosmic rays in early galaxies 
propagated
diffusively, the limits on the fieldstrength are more stringent.  We derived 
lower bounds on the fieldstrengths necessary for diffusive shock acceleration
and for confinement by 
assuming that the scattering frequency is the gyrofrequency. This corresponds
to a nonlinear level of turbulence [$B_1/B_0\sim 1$; eqn. (\ref{equ:nu})]. 
We used the upper bound
on the acceleration rate derived by \citet{LC83}
to estimate the minimum
magnetic fieldstrength required to accelerate protons to relativistic
energies [eqns. (\ref{equ:Eb}), (\ref{equ:Es}), (\ref{equ:Esb}), and
(\ref{equ:Ed})] in supernova and superbubble driven shocks, and found it to
be about 10$^{-10}$ G. This turns out to be similar to the
minimum fieldstrength required to confine cosmic rays to superbubbles long
enough for them to undergo nuclear collisions with ambient material and
synthesize the light elements observed in the oldest stars
[eqn. (\ref{equ:bubbleconf})]. At this fieldstrength, the growth rate of the
streaming instability is fast enough that the cosmic rays may be able to trap
themselves in the vicinity of the shock front [see discussion following eqn.
(\ref{equ:rootsnumerical})].

The magnetic fieldstrengths in regions of active star formation may not
have been
representative of the global galactic fieldstrength (see the discussion in
\S\ref{sec:early}). Therefore we 
derived a relationship between the average interstellar energy
density in cosmic rays, the power in cosmic ray sources, and the 
global confinement
time, assuming the maximum scattering rate. If the intensity of cosmic ray
sources and the energy density in cosmic rays were comparable to 
their current values, the magnetic field could have been as low as
10$^{-12}$ -
10$^{-13}$ G. The fieldstrength scales inversely with the luminosity of the
sources, so an enhanced supernova rate would have permitted an even smaller 
magnetic field. A field of this strength is consistent with growth of the
waves on a timescale less than the nominal confinement time of 10$^{7}$ yr,
provided that the drift velocity $v_D$ exceeds $\sim$100 km s$^{-1}$.

Estimates of the fieldstrength based on cosmic ray diffusion theory scale
with the turbulent amplitude as $(B_1/B_0)^{-2}$ [see eqn. (\ref{equ:nu})].
There are two reasons why
the turbulence could have been much stronger at early times than it
is today. The sources extraneous to
cosmic rays range from stellar radiative and mechanical
luminosity to large scale dynamical effects associated with differential
rotation and self gravity 
[\citet{SB99}, \citet{KO00}]. 
Under present conditions, the
gyroradius of a GeV cosmic ray is about 1 AU [see eqn. (\ref{equ:k0})], much
smaller than the scale at which turbulent energy is injected. But at 
fieldstrengths of 10$^{-12}$ G, the gyroradii were at parsec scales, closer to
the energy injection scale, at which the turbulent amplitude is significantly
larger.

The strength of these turbulent sources, however, can only be speculated upon.
Turbulence is driven by the anisotropy of the cosmic rays themselves.
The anisotropy is an inevitable consequence of the discrete and inhomogeneous
distribution of cosmic ray sources, and the finite size of the galaxy.
In \S\ref{sec:highbeta} we
considered the propagation, excitation, and damping of low frequency
electromagnetic waves in a medium consisting of cosmic
rays, thermal plasma, and a weak magnetic field. We found that the circularly
polarized, parallel propagating Alfv\'en waves which efficiently scatter
cosmic rays in contemporary galaxies, and which are excited by cosmic ray
streaming anisotropy, are dramatically modified
because the electrons $\bE
\times\bB$ drift in the wave electromagnetic fields but the cosmic rays do not
[Appendix A and eqn. (\ref{equ:Xdef})]. These modifications are only
significant when the pressure in the mean field is much less than the cosmic
ray pressure [see eqn. (\ref{equ:Xdef})]. 
We found that if
the drift velocity $v_D$ 
is far above the instability threshold $\vai$, 
the real and imaginary parts of the wave frequency are
comparable, and much larger than the Alfv\'en frequency [see eqn. 
(\ref{equ:rootslimitnew})]. In contrast to the current situation,
neither Landau
damping in fully ionized regions (\S\ref{subsec:landau} and Appendix B)
nor frictional damping in weakly ionized
regions (\S\ref{subsec:friction} and Appendix C)
appears capable of competing with the excitation rate.
Therefore, we
predict that the waves grow to nonlinear amplitudes, and the cosmic rays are
scattered at nearly the maximum possible rate. The limits on $B$
obtained under the assumption of scattering at the maximum rate
are probably close to correct. 

Cosmic
rays will interact dynamically with the thermal gas as long as they are
strongly scattered
[see eqns. (\ref{equ:Fcr}) and (\ref{equ:dotEcr})]. The rate of energy
transfer is usually assumed to be proportional to the Alfv\'en speed under
standard conditions, but is much larger in the
present case because the turbulence which
scatters the cosmic rays is highly super-Alfv\'enic [see eqn. 
(\ref{equ:vphivai})]. Cosmic rays could potentially play an important role in
driving outflows from young galaxies.

The fact remains that
magnetic fields and cosmic rays are in approximate equipartition
in our own Galaxy at the present time. 
We suggest that this is so because they
share a common energy source: supernovae. 
Given enough time, magnetic fields, cosmic rays, and for that matter turbulent
bulk motions
will probably achieve a state near equipartition, as currently
observed in the Galaxy. However, as we have shown
in this paper, this is not required by
anything fundamental to the physics of cosmic ray
acceleration or confinement.
Why the magnetic field and cosmic ray energy density have
saturated at values close to the turbulent energy density,
instead of increasing beyond it, is not completely clear. The reason may be
that if the
magnetic field or cosmic ray energy density were appreciably larger, the
vertical stratification of the interstellar medium would be unstable to
buoyancy driven
instabilities \citep{P66}, 
which would probably lead to separation of the thermal
and nonthermal components of the interstellar medium, and possibly facilitiate
magnetic field and cosmic ray escape.  We have briefly considered the role of
these buoyancy instabilities 
at early times, and find it to be ambiguous. If the ratio of cosmic ray
pressure to thermal gas pressure is denoted by $\alpha$, the effective
polytropic exponents of the thermal gas and cosmic rays are $\gamma_g$ and
$\gamma_{cr}$, and magnetic pressure is negligible,
 then the system is unstable if $\gamma_g -1 + \alpha(\gamma_{cr}
-1) < 0$ \citep{ZK75}. Thus, if the cosmic rays are well coupled by scattering
and $\gamma_{cr}=4/3$, they can stabilize the system, but if they diffuse
rapidly, they are destabilizing. While a detailed investigation of this 
problem is
beyond the scope of this paper, it is safe to say that if the instability
occurs, its nonlinear development differs from that of the instability in a
strong magnetic field.

If cosmic rays can be accelerated and
confined in the presence of magnetic fields which are several orders of
magnitude weaker than the fields in contemporary galaxies then evidence
for cosmic rays in young galaxies is not evidence for magnetic fields at the
equipartition level. The fieldstrengths required 
are several orders
of magnitude larger than the seed fields produced by mechanisms operating on
large scales, but
are representative of the fields expected from
a superposition of many small sources, such as plerion supernova remnants. The
dominance of small scale fields predicted by some dynamo theories is
inconsistent with the presence of cosmic rays, unless the field is strong
enough that the cosmic ray gyroradius is less than the size of the system.
Thus, the lower bounds on $B$ implied by the presence of cosmic rays 
impose meaningful constraints on theories of the
origin and evolution of galactic magnetic fields (\S\ref{sec:early}). 

Finally, the drastic changes in the properties of hydromagnetic turbulence
brought about by cosmic rays may be of some importance beyond the present
application, although they fall outside its immediate domain. 
It has been pointed out elsewhere that cosmic rays can be a
significant source of hydromagnetic fluctuations at the scale of their
gyroradius, and that this may enhance the rate at which they are accelerated
in shocks [\citet{E85}, \citet{LB00}, \citet{BL01}]. Because the  
gyroradius scale falls
between the global galactic scale and the resistive scale, fluctuations
driven by cosmic rays may affect the operation of the galactic dynamo. The
novel dispersion relation of these fluctuations will affect their nonlinear 
behavior, and will modify the properties of a turbulent cascade. These effects 
could operate in any environment
 in which the magnetic field is weak and energetic
particles are present. 

\acknowledgements

This paper was begun at the conference ``Making Light of Gravity" held in honor
of Sir Martin Rees. I am grateful to the organizers for the opportunity to
speak there. Fabian Heitsch, J. Michael Shull, and
especially an anonymous referee provided useful comments on the manuscript.
Material support was provided by NSF Grant 0098701 to
the University of Colorado.

\section*{Appendix A. Dispersion Relation with Cosmic Rays}
\begin{appendix}
\renewcommand\theequation{A\arabic{equation}}

The dispersion relation for plane electromagnetic waves in a plasma follows
from the equation for the first order electric field $\bE_1$
\begin{equation}\label{equ:gendr}
\bk\times\left(\bk\times\bE_1\right) +\f{\omega^2}{c^2}\bK\cdot\bE_1 = 0,
\end{equation}
where the response of the plasma is contained in the dielectric tensor $\bK$.
Equation (\ref{equ:gendr}) and general expressions for the components of $\bK$
are given in texts such as \citet{KT73}.

We
consider circularly polarized waves propagating parallel to the ambient
magnetic field $\hat zB_0$. In this case, $K_{yy}=K_{xx}$ and $K_{yx}=-K_{xy}$,
and  eqn. (\ref{equ:gendr}) reduces to
\begin{equation}\label{equ:pardr}
\left[K_{xx}\pm iK_{xy} -\f{c^2k^2}{\omega^2}\right]\left(E_{1x}\mp iE_{1y}
\right)=0.
\end{equation}
The $\pm$ and $\mp$ signs denote the sign of circular polarization. The
dispersion relation follows by setting the term in square brackets equal to
zero.

The dielectric terms are
\begin{equation}\label{equ:dielectric}
K_{xx}\pm iK_{xy}=1+\sum_{\alpha}\f{2\pi q_{\alpha}^2}{\omega}
\int_{0}^{\infty}\int_{-1}^{1}
\f{p^2v\left(1-\mu^2\right)}{\omega-kv\mu\pm\omega_{c\alpha}}
\left[\f{\partial f_{0\alpha}}{\partial p} + \left(\f{kv}{\omega} - \mu\right)
\f{1}{p}\f{\partial f_{0\alpha}}{\partial\mu}\right]dpd\mu,
\end{equation}
where the summation index $\alpha$ represents particle species and the $f_{0
\alpha}$
are the zero order phase space distribution functions averaged over gyrophase.
The $\omega_{c\alpha}$ are the relativistic gyrofrequencies.

We evaluate eqn. (\ref{equ:dielectric}) for
a system with four populations of particles:
cold (zero temperature) protons with number density $n_i$
drifting with speed $-\hat z v_D$, cold electrons with the same number density
and drift speed as the ions, proton cosmic rays which are isotropic in
velocity space and have number density $n_{cr}$, and an
isotropic population of cold electrons, also with number density $n_{cr}$. This
choice guarantees that to zero order the system is electrically neutral and
there is no net current flow, consistent with the assumption of a uniform
magnetic field. We have chosen to work in the frame in which the
cosmic rays are isotropic, and transform the
resulting dispersion relation to the rest frame of the
plasma, because that is the simplest way to do the calculation.

\citet{FK79} considered the effect of finite temperature (but did
not include cosmic rays). As we argue in \S\ref{subsec:landau}, Landau 
damping is the most
important thermal effect. It requires wave propagation at an angle to $\bB_0$,
and is considered in Appendix B.

The distribution functions of the cold, drifting species are
\begin{equation}\label{equ:colddrift}
f_{0\alpha}=\f{n_i}{p^2}\delta\left(p-p_{D\alpha}\right)\delta\left(\mu + 1
\right),
\end{equation}
where $p_{D\alpha}\equiv m_{\alpha}v_D$, and $\alpha$ represents protons or
electrons. The distribution function of the cold, stationary electrons is
\begin{equation}\label{equ:coldstatry}
f_{0\alpha}=\f{1}{2}\f{n_{cr}}{p^2}\delta(p),
\end{equation}
and we write the distribution function of the cosmic rays as
\begin{equation}\label{equ:cr}
f_{0cr}=\f{1}{2}n_{cr}\phi(p),
\end{equation}
where $\phi(p)$ is normalized to unity. Substituting eqns. (\ref{equ:colddrift}), (\ref{equ:coldstatry}), and (\ref{equ:cr}) into eqn. (\ref{equ:dielectric})
and carrying out the integrations gives the dispersion relation
\begin{eqnarray}\label{equ:dr1}
\f{c^2k^2}{\omega^2}&=&1\mp\f{4\pi n_i e^2}{\omega^2}\left(\omega + kv_D\right)
\left[\f{1}{m_e}\f{1}{(\omega_{ce}\pm\omega\pm kv_D)} +
\f{1}{m_i}\f{1}{(\omega_{cp}\pm\omega\pm kv_D)}\right]\nonumber \\
&  &\mp\f{4\pi n_{cr} e^2}{\omega}
\left[\f{1}{m_e}\f{1}{(\omega_{ce}\pm\omega)}
\pm\f{i\pi}{4}\f{p_1}{m_p\omega_{cp}}\int_{p_1}^{\infty}\left(p^2-p_1^2\right)
\f{d\phi}{dp}dp\right.\nonumber \\
&  &\left. \pm\f{1}{m_p\omega_{cp}}\f{\mathcal{P}}{4}\int_{0}^{\infty}\int_{-1}^{1}
\f{p_1p^2\left(1-\mu^2\right)}{\mu\mp p_1/p}\f{d\phi}{dp}dpd\mu\right],
\end{eqnarray}
where ${\mathcal{P}}$ denotes the principal part of the integral, 
\begin{equation}\label{equ:p1} 
p_1\equiv\f{ m_p\omega_{cp}}{k} 
\end{equation}
is the minimum momentum at which the
resonance condition eqn. (\ref{equ:resonance}) can be satisfied, and we have
dropped $\omega$ compared with the other terms in the denominator of the last
integrals. The right hand side of eqn. (\ref{equ:dr1}) is $K_{xx}\pm
iK_{xy}$.

The ``1" on the right hand side of eqn. (\ref{equ:dr1}) represents the
displacement current. The first square bracket represents the thermal plasma.
Since the gyrofrequencies $\omega_{cp}$, $\omega_{ce}$ much exceed $\omega$
and $kv_D$, we approximate the electron term as $1/m_e\omega_{ce}$ and the
proton term as $1/m_p\omega_{cp}\mp (\omega+kv_D)/m_p\omega_{cp}^2$. The
dominant terms cancel: $\omega_{ce}m_e=-\omega_{cp}m_p$. The cancellation
occurs because in the low frequency waves considered here, the electron and
ion motion is nearly the $\bE\times\bB$ drift, which gives no net current.

The second square bracket represents the cosmic rays and the cold electrons
that cancel their current. The first cosmic ray integral
represents the resonant cosmic rays. It is this term which leads to instability
if the relative drift of the ions and thermal plasma exceeds $\vai$. The
second term represents the nonresonant cosmic rays.
If we revert to the cold
plasma approximation by taking the limit $p_1/p\gg 1$, this term reduces to
$1/m_p\omega_{cp}$, which cancels the electron term. The condition $p_1/p\gg 1$
is equivalent to the condition that the wavelength of the wave is much larger
than the particle gyroradius. If this condition is not fulfilled, the particles
do not $\bE\times\bB$ drift, and their current does not cancel the cold 
electron current. 
This can lead to a relatively large
effect on the dispersion relation even when $n_{cr}/n_i \ll 1$.

We simplify the cosmic ray term somewhat by integrating the resonant term by
parts and performing the integration over $\mu$ in the nonresonant term. The
cosmic ray integrals in the second square bracket of eqn. (\ref{equ:dr1}) then
reduce to
\begin{equation}\label{equ:crterm}
\mp\f{i\pi}{2}\f{p_1}{m_p\omega_{cp}}\int_{p_1}^{\infty}p\phi dp
\pm\f{p_1}{m_p\omega_{cp}}\f{\mathcal{P}}{4}\int_{0}^{\infty}\left[\left(p^2
-p_1^2\right)\ln{\left\vert\f{1\mp p/p_1}{1\pm p/p_1}\right\vert}\mp 2pp_1
\right]\f{d\phi}{dp}dp.
\end{equation}
We then
incorporate the electrons directly into the nonresonant cosmic ray term by
replacing $1/m_e(\omega_{ce}\pm\omega)$ by $-1/m_p\omega_{cp}$.

Finally, we drop the displacement current and multiply eqn. (\ref{equ:dr1}) by
$\omega^2\vai^2/c^2$. The resulting dispersion relation is
\begin{eqnarray}\label{equ:dr2}
k^2\vai^2&=&\left(\omega + kv_D\right)^2 + 
\omega\omega_{cp}\f{n_{cr}}{n_i}\times \nonumber \\
&  &\left\{\f{i\pi}{2}\int_{p_1}^{\infty}p_1p\phi dp - \f{p_1\mathcal{P}}{4}
\int_{0}^{\infty}\left[\left(p^2
-p_1^2\right)\ln{\left\vert\f{1\mp p/p_1}{1\pm p/p_1}\right\vert}\mp 2pp_1
\pm\f{4}{3}\f{p^3}{p_1}\right]\f{d\phi}{dp}dp\right\}.
\end{eqnarray}
If $v_D\equiv 0$ and
 $p/p_1$ is small but finite for the bulk of the particles, eqn. (\ref{equ:dr2}) reverts to the dispersion relation for hydromagnetic waves in a hot plasma
\citep{FK79}, with a factor of $n_{cr}/n_i$ multiplying the thermal correction
term.

\begin{figure}[h]
\epsscale{.7}
\plotone{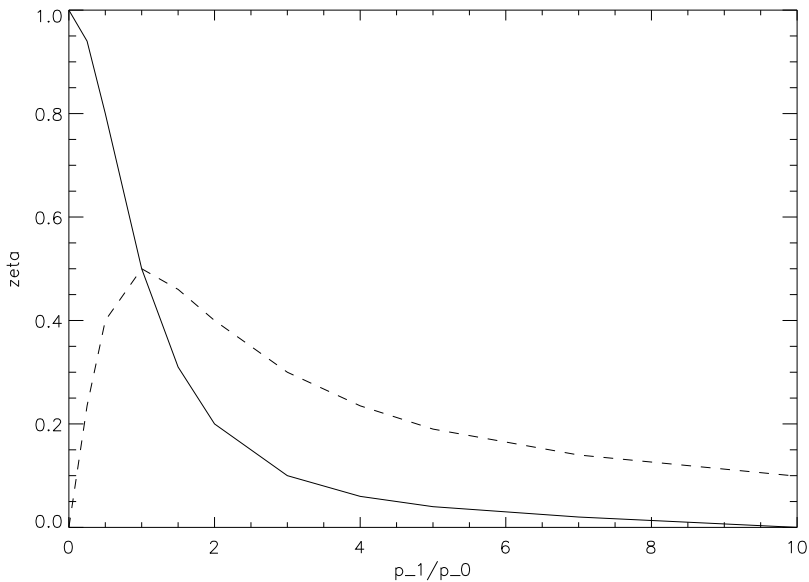}
\caption{\label{fig:zeta} Absolute value of $Re(\zeta)$ (solid curve) and 
$Im(\zeta)$ (dashed curve) \textit{vs} $p_1$ for the distribution function
given in eqn. (\ref{equ:exphi}).}
\end{figure}

The quantity in $\{\}$ is the function $\zetalr$ introduced in eqn.
(\ref{equ:epscrdef}). It is clear from eqn. (\ref{equ:dr2}) that $\zeta_l=-
\zeta_r^{\ast}$. The behavior of $\zeta$ is illustrated in Figure 
\ref{fig:zeta} for the normalized distribution function
\begin{equation}\label{equ:exphi}
\phi(p)=\f{4}{\pi p_0^3}\left(1+\f{p^2}{p_0^2}\right)^{-2},
\end{equation}
which is close to the Galactic cosmic ray spectrum for $p/p_0\gg 1$ (the
observed power
law index is approximately 4.6 rather than 4).
Equation (\ref{equ:dispreli}) follows from eqn. (\ref{equ:dr2}) by 
shifting back to the rest frame of the thermal plasma.
\end{appendix}

\section*{Appendix B. Landau Damping Rate}
\begin{appendix}
\renewcommand\theequation{B\arabic{equation}}

We calculate the Landau damping rate using the dispersion relation derived by
\citet{FK79}, but replacing the thermal terms entering into the real part of
$\omega$ with cosmic ray terms. The dispersion relation for small propagation
angle $\theta$ can be written as
\begin{equation}\label{equ:drtheta}
\left(K_{xx} + iK_{xy} - \f{c^2k^2}{\omega^2}\right)\left(K_{xx} - iK_{xy} -
\f{c^2k^2}{\omega^2}\right)+i\pi^{1/2}\f{kv_i}{\omega}\left(K_{xx}-
\f{c^2k^2}{\omega^2}\right)\f{c^2}{\vai^2}\tan^2{\theta}=0,
\end{equation}
where $v_i\equiv (2k_BT/m_p)^{1/2}$, and the components of the dielectric
tensor are given in Appendix A. For $\theta=0$, eqn. (\ref{equ:drtheta})
reduces to the dispersion relation eqn. (\ref{equ:pardr}).

We write $\omega=\omega_0 + \epsilon\omega_1$, where $\omega_0$ is a solution
of eqn. (\ref{equ:drtheta}) for $\theta = 0$ and assume $\epsilon\sim
{\mathcal{O}}(\theta^2)\ll 1$. Expanding eqn. (\ref{equ:drtheta}) to first order
in $\epsilon$ and using the dispersion relation at $\theta=0$ gives
\begin{equation}\label{equ:omega1}
\epsilon\omega_1=-i\f{\pi^{1/2}}{2}\f{kv_i}{\omega}
\f{c^2}{\vai^2}{\mathcal{D}}(\omega_0)^{-1}\tan^2{\theta}
\end{equation}
where
\begin{equation}\label{equ:D}
D(\omega_0)\equiv\f{\partial}{\partial\omega}\left(K_{xx}\pm iK_{xy}\right)
\vert_{\omega_0}.
\end{equation}
If the waves are Alfv\'en waves, $\omega_0^2=k^2\vai^2$, $D=2c^2k^2/\omega^3$,
and
\begin{equation}\label{equ:damping1}
\epsilon\omega_1=i\f{\pi^{1/2}}{2}\f{kv_i}{\omega}\tan^2{\theta},
\end{equation}
which to order unity is the result of \citet{FK79}. If the parameter $X$
defined in eqn. (\ref{equ:Xdef}) is large, $\omega\sim(\epscr\omega_{cp}kv_D)^{1/2}$ [see eqn. (\ref{equ:rootslimitnew})], and $D\sim 2c^2/\omega\vai^2$.
Substituting this result into eqn. (\ref{equ:omega1}) shows that $\epsilon\omega_1$ is still given by eqn. (\ref{equ:damping1}).

\citet{FK79} found that the waves are critically damped - the real and
imaginary parts of $\omega$ are comparable - when $\theta\ge(\vai/v_i)^{1/2}$.
When the wave frequency is determined by cosmic rays, the angle for critical
damping is increased to $\sim(cv_D\epscr/v_i^2)^{1/4}\sim 10(v_D/c)^{1/4}
(n_iT)^{-1/4}$. The damping is weakened because the thermal plasma energy is
a relatively small part of the wave energy, as shown in Appendix C.
\end{appendix}

\section*{Appendix C. Damping Rate from Ion-Neutral Friction}
\begin{appendix}
\renewcommand\theequation{C\arabic{equation}}
We calculate the damping rate $\Gamma_{in}$ using an energy method, rather than
directly solving the dispersion relation 
%\citep{B65}. 
Let the energy density
in waves be $U_w$ and the rate per unit volume at which energy is dissipated be
$\dot U_w$. These quantities are related to $\Gamma_{in}$ by
\begin{equation}\label{equ:basic}
\dot U_w = -2\Gamma_{in}U_w.
\end{equation}
The dissipation rate can be written in terms of the first order thermal ion
and neutral velocities $\bvv_{1i}$ and $\bvv_{1n}$ as
\begin{equation}\label{equ:uwdot}
\dot U_w=-\rho_i\nu_{in}\mid\bvv_{1i}-\bvv_{1n}\mid^2.
\end{equation}
The wave energy density includes contributions from the electric and magnetic
fields, thermal plasma, and cosmic rays (the contribution from the neutrals is
much smaller than the plasma contribution because the wave frequency is so
high that the neutrals remain nearly at rest), and can be written as 
\citep{B66}
\begin{equation}\label{equ:uw}
U_w=\f{1}{8\pi}\bB_1^{\ast}\cdot\bB_1 + \f{1}{8\pi}\bE_{1}^{\ast}\cdot
\f{\partial}{\partial\omega}\left(\omega\bK\right)\cdot\bE_1,
\end{equation}
where $\bK$ is the dielectric tensor (see Appendix A).
The first term in eqn. (\ref{equ:uw}) represents magnetic energy density, and
the second accounts for the energy density in the electric field, thermal
plasma, and cosmic rays.

We assume the waves are circularly polarized and
propagate parallel to $\bB_0$, which we take to define the
$\hat z$ direction, and that first order quantities vary with $z$ and $t$ as $
\exp{i(\omega t - kz})$.
The neutral velocity is determined solely by friction
with the ions, and can be written in terms of $\bvv_{1i}$ as
\begin{equation}\label{equ:neutrals}
i\omega\bvv_{1i}=\nu_{ni}\left(\bvv_{1n}-\bvv_{1i}\right),
\end{equation}
where the neutral-ion collision frequency $\nu_{ni}=\rho_i\langle\sigma v
\rangle/(m_i+m_n)$. The ion velocity is determined by both frictional and
magnetic forces (it suffices to use a one fluid treatment of the plasma,
$\bvv_{1i}\approx\bvv_{1e}$, because the thermal electron and ion velocities
are 
essentially the $\bB\times\bB$ drift. The
ion equation of motion is
\begin{equation}\label{equ:ions}
i\omega\bvv_{1i}=-\f{i}{4\pi\rho_i}kB_0\bB_1 + \nu_{in}\left(\bvv_{1n}-\bvv_{1i}\right).
\end{equation}
Solving eqn. (\ref{equ:neutrals}) for $\bvv_{1n}$ in terms of $\bvv_{1i}$ and
expressing $\bvv_{1i}$ in terms of $\bB_1$, we rewrite eqn. (\ref{equ:uwdot})
as
\begin{equation}\label{equ:uwdot2}
\dot U_w=-\nu_{in}\f{\vai^2}{v_{\phi}^2}\f{1}{4\pi}\bB_1^{\ast}\cdot\bB_1,
\end{equation}
where we have assumed $\omega/\nu_{in}\gg 1$.

The second term on the right hand side of eqn. (\ref{equ:uw}) can be written as
\begin{equation}\label{equ:epenergy}
\f{1}{4\pi}E_{1x}E_{1x}^{\ast}\f{\partial}{\partial\omega}\omega\left(K_{xx}\pm
iK_{xy}\right).
\end{equation}
From eqns. (\ref{equ:pardr}) and (\ref{equ:rootslimitnew}),
\begin{equation}\label{equ:KxxKxy}
K_{xx}\pm iK_{xy}\approx -\epscr\f{c^2}{\vai^2}\f{\omega_{cp}kv_D}{\omega^2}.
\end{equation}
It is straightforward to evaluate eqn. (\ref{equ:epenergy}) using eqn.
(\ref{equ:KxxKxy}). We replace
$E_{1x}E_{1x}^{\ast}$ by $\bB_1
\cdot\bB_1^{\ast}v_{\phi}^2/2c^2$ using Faraday's law. Equation (\ref{equ:uw})
becomes
\begin{equation}\label{equ:uw2}
U_w=\f{1}{8\pi}\bB_1\cdot\bB_1^{\ast}\left(1+\f{v_{\phi}^2}{\vai^2}\right).
\end{equation}
The first term in eqn. (\ref{equ:uw2}) represents the magnetic energy density,
while the second is primarily cosmic ray energy, with a small contribution
from the first order electric field.

Equation
(\ref{equ:friction})
follows from eqn. (\ref{equ:basic}) if we substitute for $U_w$ and
$\dot U_w$ from eqns. (\ref{equ:uwdot2}) and (\ref{equ:uw2}).

\end{appendix}

%%%%%%%%%%%%%%%%%%%%%%%%%%%%%%%%%%%%%%%%%%%%%%%%%%%%%%%%%%%%
%
% section{Bibliography}
%
%%%%%%%%%%%%%%%%%%%%%%%%%%%%%%%%%%%%%%%%%%%%%%%%%%%%%%%%%%%%

\end{document}